\documentclass[a4paper,superscriptaddress,twocolumn,preprintnumbers,showpacs,amsmath,amssymb,prl]{revtex4}
\usepackage{graphicx}
\usepackage{latexsym}
\usepackage{mathrsfs} 

\usepackage[greek,frenchb]{babel}
\usepackage[latin1]{inputenc}

\begin{document}
\preprint{submitted for publication in \textit{Phys. Rev. Lett.}}

\title{Pink noise of ionic conductance through single artificial nanopores revisited}

\author{C. Tasserit}
\affiliation{Laboratoire L\'eon Brillouin, CEA/CNRS UMR 12, CEA-Saclay, 91191 Gif-sur-Yvette cedex, France}
\affiliation{Laboratoire des Solides Irradi\'es, CEA/CNRS/Ecole Polytechnique, Ecole Polytechnique, 91128 Palaiseau cedex, France}

\author{A. Koutsioubas}
\author{D. Lairez}
\affiliation{Laboratoire L\'eon Brillouin, CEA/CNRS UMR 12, CEA-Saclay, 91191 Gif-sur-Yvette cedex, France}

\author{G. Zalczer}
\affiliation{Service de Physique de l'Etat Condens\'{e}, CEA-Saclay, 91191 Gif-sur-Yvette cedex, France.}

\author{M.-C. Clochard}
\affiliation{Laboratoire des Solides Irradi\'es, CEA/CNRS/Ecole Polytechnique, Ecole Polytechnique, 91128 Palaiseau cedex, France}

\date{\today} 

\begin{abstract}
We report voltage-clamp measurements through single conical nanopore obtained by chemical etching of a single ion-track in polyimide film. Special attention is paid on the pink noise of the ionic current (i.e. $1/f$ noise) measured with different filling liquids. The relative pink noise amplitude is almost independent of concentration and pH for KCl solutions, but varies strongly using ionic liquids. In particular we show that depending on the ionic liquid, the transport of charge carriers is strongly facilitated (low noise and higher conductivity than in the bulk) or jammed. These results show that the origin of the pink noise can be ascribed neither to fluctuations of the pore geometry nor to the pore wall charges but rather to a cooperative effect on ions motion in confined geometry.

\end{abstract}
 

\maketitle

After the work of the Nobel prize winners E.~Neher and B.~Sakmann for their single ion-channel recordings experiments~\cite{NEHER:1976uq}, the first application of their technics for \textit{in vitro} single-molecule manipulation~\cite{Kasianowicz:1996} stimulated many hopes for the study of biological macromolecules. The main idea is that individual polymer chain driven electrophoretically through a single nanopore (namely chain translocation) causes a resistive pulse of the ionic conductance of the channel that can be observed. Noticeable challenging applications of this method are DNA sequencing~\cite{Akeson:1999, Meller:2001,Nakane:2003} and protein folding-unfolding studies~\cite{Oukhaled:2007}. Initially concerned with biological nanopores (mainly $\alpha$-hemolysin inserted into lipid bilayer), more recent reports consider artificial nanopores because of their versatility~\cite{Howorka:2009}. Two main processes are used to obtain such artificial nanopores: chemical etching of a single ion-track in polymer film~\cite{Siwy:2002fk,Mara:2004qy,Harrell:2006} and ion-beam sculpting of silicon nitride~\cite{Li:2001vn}. Nanopore sensing of macromolecules in solution is based on an accurate analysis of the electrical ionic current through the nanopore: passing throughout the nanopore, a macromolecule causes fluctuations of the ionic current. Analysis of the time correlation of these fluctuations, i.e. duration and frequency probabilities, is expected to sign the solute~\cite{Howorka:2009}. While quite promising, progress in this domain is widely impeded by a low frequency $1/f$ noise (named "pink noise") of the power spectral density (PSD) of current observed even for a nanopore filled with a blank sample i.e. solvent alone. The understanding and reduction of this noise is crucial to make the most of translocation studies~\cite{Tabard:2007yf,Smeets:2007}. Pink noise of the PSD is characteristic of anomalous and slow relaxation of fluctuations. Unfortunately, it is not the signature of a unique and universal elementary mechanism, as many causes can result in the same $1/f$ spectrum~\cite{Weissman:1988rv}. Actually, it is reported in many voltage clamp studies not only on artificial nanopores but also on biological systems from neural ionic channels~\cite{Derksen:1966}, membrane-active peptides~\cite{Sauve:1978,Fadda:2009} to protein channels~\cite{Nekolla:1994,Bezrukov:2000bs}. A common feature of pink noise encountered in electronic devices but also on ionic current though nanopores, is that the amplitude of the $1/f$ power law of the PSD increases as the square of the average current~\cite{Hooge:1981}. This is the signature of conductance fluctuations. Generally speaking, the conductance $G=I/U$ (where $I$ is the current and $U$ the voltage) of a ohmic system made of a nanopore filled with an ionic solution can be written as the product: 
\begin{equation}\label{eq1}
G=Q^2 \times C \mu \times L
\end{equation}
where $Q$ is the charge of ions, $C$ their concentration (number per unit volume), $\mu$ their effective mobility along the pore axis (averaged velocity per unit force) and $L$ an effective length characteristic of the pore geometry (typically the ratio of the cross section to the length for a cylindrical pore).
Straightforwardly from Eq.\ref{eq1}, conductance fluctuations can be either imputed to the pore itself ($L$) or to the charge transport ($C\mu$). Actually in the literature, both kinds of hypothetical explanations are proposed. Among the first kind, "channel breathing" is invoked for protein channel~\cite{Bezrukov:2000bs} and "pore wall dandling fragments" or "opening-closing" process~\cite{Siwy2002} are invoked for track-etched nanopores. Whereas the second kind is mainly proposed for silicon nitride nanopores, for which conductance fluctuation are attributed to fluctuations of ion concentration~\cite{Smeets:2007} and inspired by the Hooge phenomenological formula~\cite{Hooge:1981} obtained for electronic devices. In this case, concentration fluctuations are claimed to be related to the surface charge of the pore wall~\cite{Powell:2009fk}.

In this paper we address the problem of pink noise measured on conical track-etched single nanopore in polyimide film (Kapton). First we show that for the same level of ionic current, the pink-noise amplitude is considerably decreased using an appropriate ionic liquid as charge carrier, and considerably increased using an other one. This result gives a strong evidence that the origin of the pink noise cannot be attributed to fluctuations of the pore geometry ($L$ in Eq.\ref{eq1}) but rather to local fluctuations of the liquid conductivity ($\sigma=Q^2 C \mu$). In addition, we show that these latter fluctuations are independent on the surface charge of the pore wall and cannot be accounted for by the Hooge formula and concentration fluctuations. Thus, our results give the first evidence that the pink noise in such nanopores comes from anomalous cooperative fluctuations of the confined ions motion.

\emph{Samples characteristics and preparation:}
Single heavy-ion (Kr$^{28+}$, 10.36\,MeV) irradiations of 8\,$\mu$m thick polyimide foils (Kapton HN) were performed at GANIL (France).
Conically shaped single nanopores were prepared by anisotropic chemical etching of these irradiated films. The etching process was performed following ref.~\cite{Siwy:2003fk} at $T=328$\,K using a two chambers conductivity cell where one chamber is filled with NaOCl etching solution ($\text{pH}=12.5$), while the other chamber contains  1\,M KI neutralizing solution. Across the film a voltage of +1\,V (with respect to the grounded neutralizing compartment) is applied for detection of the breakthrough event and also to assist the neutralization of NaOCl upon breakthrough. For conductivity measurements, the ionic liquids used are 1-ethyl-3-methylimidazolium thiocyanate (EMIM-SCN, from Sigma) and 1-butyl-3-methylimidazolium  bis(trifluorosulfonyl)imide (BMIM-TFSI, from Solvionic). Their main characteristics are summarized in Tab.\ref{tab1}. They have comparable viscosity and electrical conductivity but the former is fully miscible in water whereas the latter is not. These ionic liquids display large electrochemical windows that prevent electrochemical reaction at the electrodes at our working voltage~\cite{Galinski:2006,Pringle:2002}. However, due to the very low ionic current level through a single nanopore, the polarization characteristic time of the electrodes is very long compared to our measurement duration.

\begin{table}[htdp]
\caption{Viscosity $\eta$ and electrical conductivity $\sigma_{\text{bulk}}$ of ionic liquids in the bulk at room temperature (from ref.~\cite{Galinski:2006,Domanska:2010vn}). $\sigma_{\text{conf}}$ is the electrical conductivity deduced from our measurements of conductance through nanopores.}
\begin{center}
\begin{tabular}{l|c|c|c}
  & $\eta$ (mPas) & $\sigma_{\text{bulk}}$ (S/m) & $\sigma_{\text{conf}}$ (S/m)\\
\hline
EMIM-SCN  & 20 & 0.20 & $0.23\pm 0.02$\\
BMIM-TFSI & 50 & 0.38 & $4.7\pm 0.5$\\
\hline
\end{tabular}
\end{center}
\label{tab1}
\end{table}

\emph{Data acquisition and treatment:}
Voltage clamp and current amplification were ensured by an Axopatch 200B with a 10\,kHz low-pass analog filter setting. The amplified current was digitized with a 16 bits ADC (Iotech Dacqbook) at 250\,kHz sampling rate and averaged over 25 samples. PSD was averaged following the periodogram method over at least 50 time segments. Measurements were performed at room temperature under normal atmospheric composition using two Ag/Ag-Cl electrodes of 2\,mm diameter and 10\,mm in length with the tip of conical nanopore at the ground potential.

\begin{figure}[!hbtp]
\centering
\includegraphics[width=0.8\linewidth]{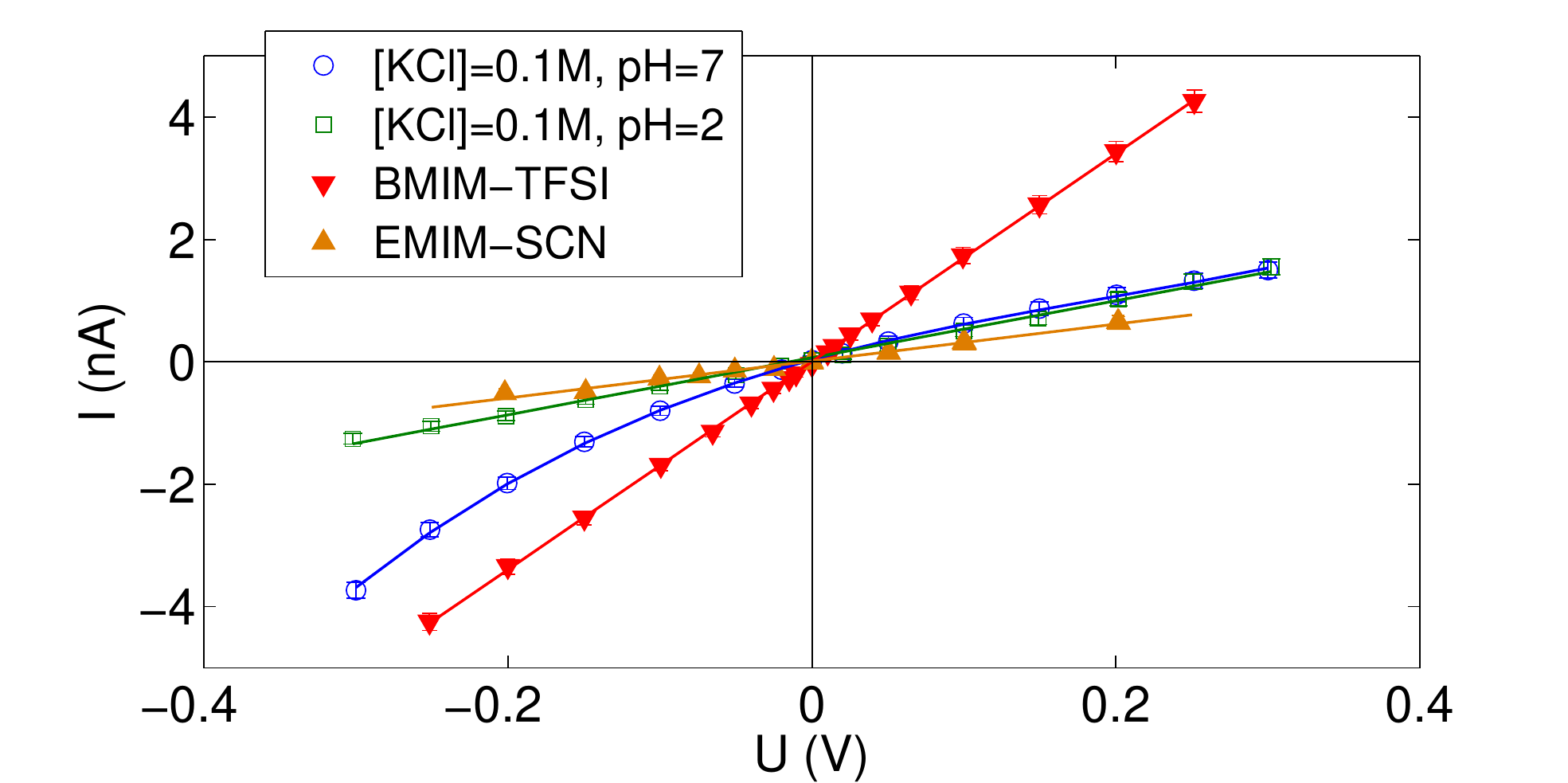}
\caption{Current-voltage characteristic curves measured for single nanopore with different filling liquids. Slopes of straight lines are equal to 3.0\,nS (EMIM-SCN), 17\,nS (BMIM-TFSI) and 4.7\,nS ([KCl]~=~0.1\,M), respectively.}
\label{figIV}
\end{figure}

\emph{Results:}
In Fig.\ref{figIV}, typical current-voltage characteristic curves of a single conical nanopore are plotted for different filling liquids. For KCl solutions (molar conductivity: 73 and $76\times 10^{-4}$\,S\,m${^2}$\,mol$^{-1}$ for K$^+$ and Cl$^-$, respectively) at pH~=~7, the nanopore is highly rectifying. As ionic conductivities of cations and anions are identical, the symmetry breaking can only be due to the electrical charge of the pore wall. Oxydation during chemical etching leads to carboxylic groups on the pore wall which are dissociated at pH~=~7. This charged surface is responsible for an ion selectivity leading to this polarity dependent conductance of the pore. At pH~=~2, carboxylic groups are fully protonated, the pore wall is neutral and this effect disappears. At this pH, the variation of the nanopore conductance with KCl concentration (up to 3\,M) does not differ significantly from the variation of KCl conductivity in the bulk reported in ref.~\cite{Monica:1984}. The effective characteristic length $L$ of the pore can thus be determined from the ratio of the conductance $G$ to the conductivity $\sigma$ of the filling solution: $L=G/\sigma$. The results here reported were obtained using two nanopores of characteristic size $L=(4. 0\pm 0.5)$\,nm. For truncated conical pores: $L=\pi r_1r_2/l$, where $r_1$ and $r_2$ are the radii of the two apertures and $l$ the pore length (film thickness). The largest radius $r_1$ has been measured by "field emission scanning electron microscopy" imaging of a multipore membrane (10$^8$ pores cm$^{-2}$) prepared under the same conditions as single pores and was found to be $r_1=0.5$\,$\mu$m. Single nanopores differ the one from the other mainly by their smaller radius $r_2$. From the $L$ values one gets $r_2=(20\pm 2)$\,nm. The conductance measured with ionic liquid compared to KCl solutions allows us to determine the conductivity $\sigma_{\text{conf}}$ of ionic liquids in the nanopore (Tab.\ref{tab1}). For EMIM-SCN (hydrophilic anion), $\sigma_{\text{conf}}\simeq \sigma_{\text{bulk}}$. However, for BMIM-TFSI (hydrophobic anion), one finds $\sigma_{\text{conf}}\gg \sigma_{\text{bulk}}$. Note that recently with similar nanopores filled with BMIM-methyl sulfate and methoxyethoxyethyl sulfate (amphiphilic anion) the opposite behavior is reported ($\sigma_{\text{conf}}\ll \sigma_{\text{bulk}}$)~\cite{Davenport:2009fk}. As regards to the wide variability of ionic liquids properties these discrepancies are not necessarily unexpected and opens a wide field of investigation.

\begin{figure}[!htbp]
\centering
\includegraphics[width=0.8\linewidth]{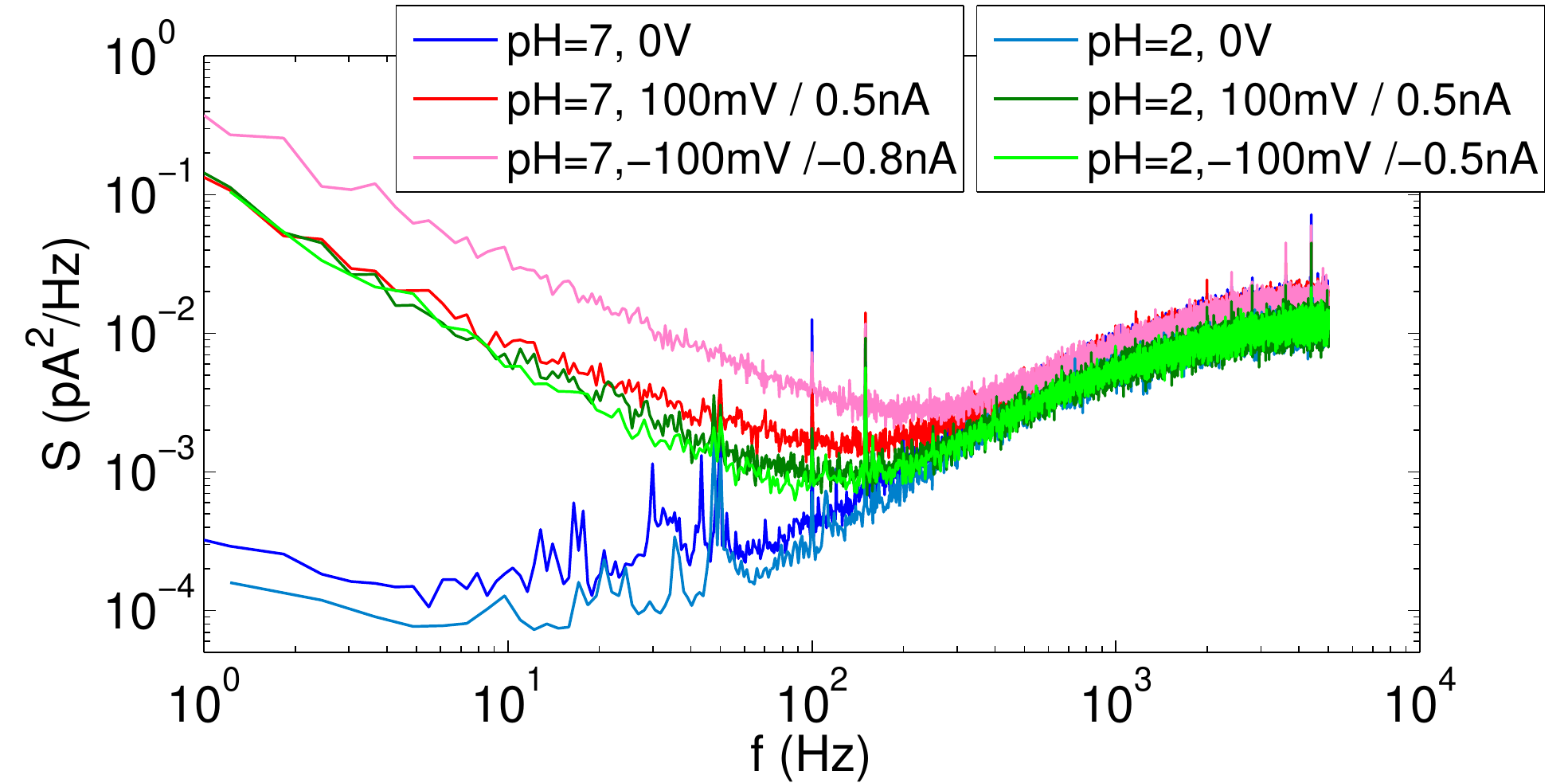}
\caption{Power spectral density, $S$, of the ionic current through a single nanopore filled with KCl solutions ([KCl]~=~0.1\,M) at different pH values and applied voltages.}
\label{figSF1}
\end{figure}
\begin{figure}[!htbp]
\centering
\includegraphics[width=0.8\linewidth]{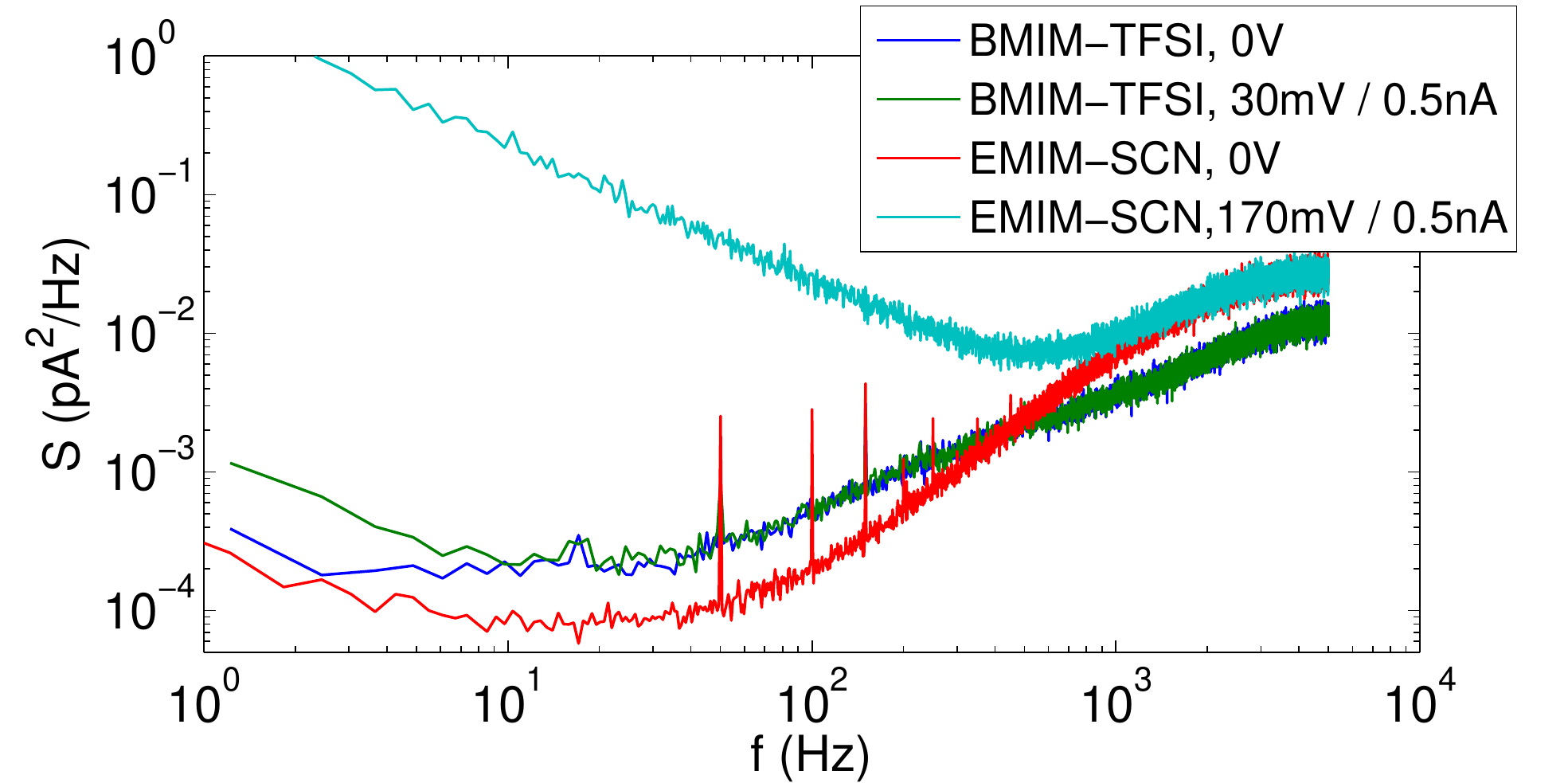}
\caption{Power spectral density, $S$, of the ionic current through single nanopore filled with ionic liquid for $I=0$ and $0.5$\,nA.}
\label{figSF2}
\end{figure}

\begin{figure}[!hbtp]
\centering
\includegraphics[width=0.8\linewidth]{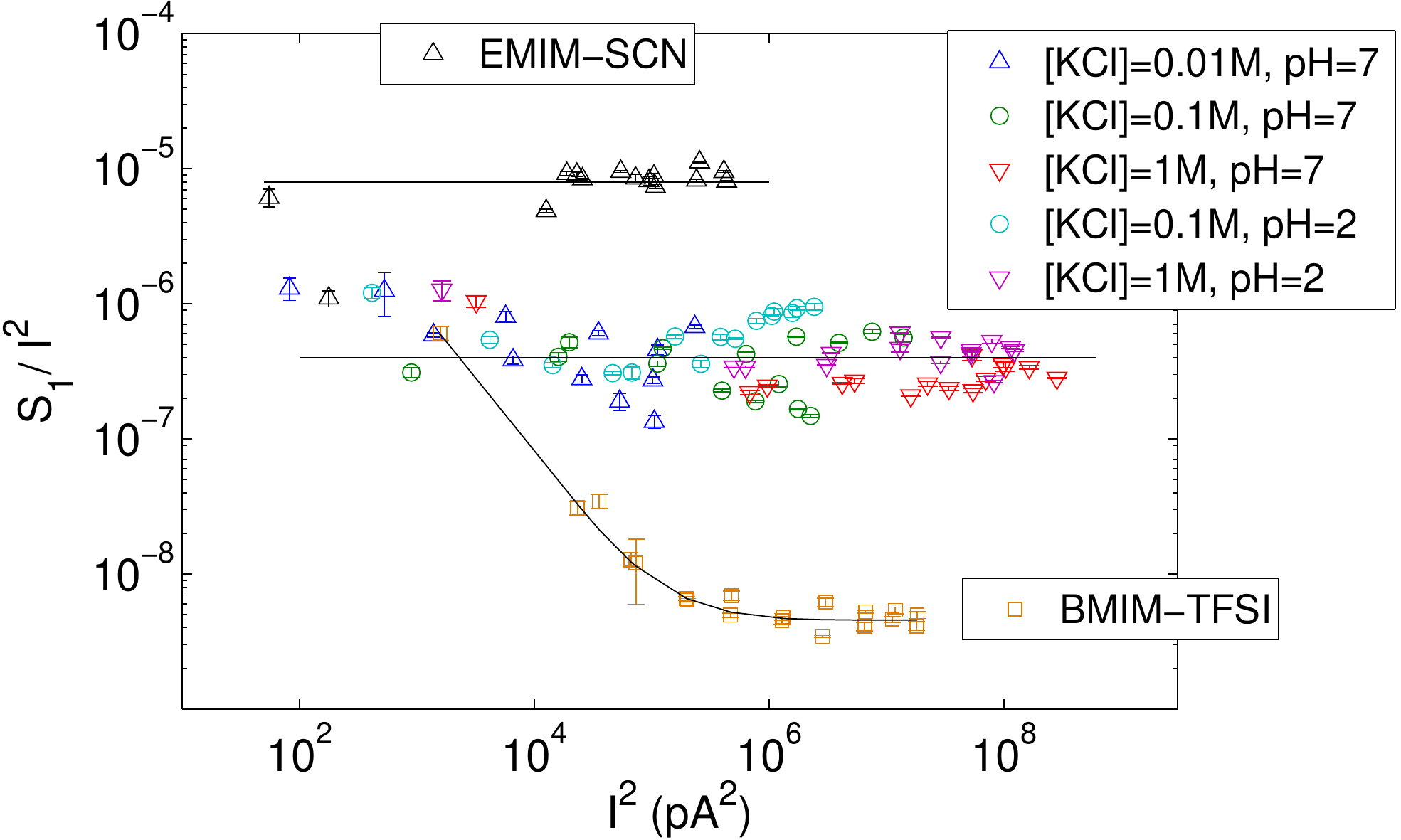}
\caption{Relative pink noise amplitude $S_1/I^2$ vs. square of the current $I^2$ for single nanopores with different filling liquids at different voltages. Lines are guides for the eyes.}
\label{figSI2}
\end{figure}

In Fig.\ref{figSF1}, typical power spectral density of the current is plotted for KCl solution at pH~=2 and 7 at different voltages. Typical spectra measured for ionic liquids are plotted in Fig.\ref{figSF2}. The shape of these spectra can be imputed neither to the electrodes nor to the measurement device~\cite{Hoogerheide:2009fk}, it displays two parts. At high frequency the spectra are independent of the voltage and of the pH. This high frequency noise can be attributed to electrochemical equilibrium of functional groups of the pore wall~\cite{Hoogerheide:2009fk} and can be fitted with a polynomial~\cite{levis:1993}. At low frequency spectra display a $1/f$ noise that increases in amplitude with the current. The whole frequency range was accounted for by fitting the spectra with:
\begin{equation}\label{eqS}
S=S_1\frac{1}{f} + a + bf - cf^2 +\cdots
\end{equation}
Results found for the pink noise amplitude $S_1$ are plotted in Fig.\ref{figSI2}. For KCl solutions, whatever the salt concentration, the pH and the voltage, a single master curve $S_1/I^2=\text{Cst}$ is found over 6 orders of magnitude. With ionic liquids, the amplitude of the pink noise differs significantly from the KCl solutions master curve. It is increased by a factor 40 with EMIM-SCN but conversely decreased by 2 orders of magnitude with BMIM-TFSI. The origin of this discrepancy is not understood but should be probably related to the effect of confinement on the conductivity already mentioned.

\emph{Discussion:}
The power spectral density $S$ is the Fourier transform of the autocorrelation that obviously vanishes beyond the longest relaxation time $\tau^*=1/f^*$. In our case one can write
\begin{equation}\label{eqS}
S=t_0 I^2\frac{\left<\Delta G^2\right>}{G^2} \mathcal{F}(f/f^*) \text{, with}\left\{\begin{array}{l}\mathcal{F}(x\ll 1)=1\\\mathcal{F}(x\gg 1)=x^{-1}\end{array}\right.
\end{equation}
with $t_0$ the time unit. Unfortunately, we did not succeed in reaching the expected plateau of the PSD at low frequency. 

Let us first consider fluctuations of the pore geometry (breathing or dandling fragments) as responsible for the pink noise. $G=\sigma L$, $\left<\Delta G^2\right>=\sigma^2\left<\Delta L^2\right>$ and Eq.\ref{eqS} give: 
\begin{equation}
S_1\propto I^2\frac{\left<\Delta L^2\right>}{L^2}f^*
\end{equation}
One can reasonably assume that the amplitude $\left<\Delta L^2\right>$ of fluctuations of pore geometry is a thermodynamical or static property that only weakly depends on the filling liquid. Only the dynamics of these fluctuations would shift to low frequencies proportionally to the increase of viscosity $\eta$, i.e. $f^*\propto \eta^{-1}$. Finally, fluctuations of the pore geometry would lead to $S_1\propto I^2 \eta^{-1}$. Evidently our results disagree with this behavior as both ionic liquids have a higher viscosity than KCl solutions but give a pink noise much higher or much lower.

The pink noise is more likely to come from fluctuations of ionic conductivity of the confined liquid, i.e. concentration or mobility fluctuations. Let us consider $N$ independent charge carriers with individual current contributions $i$: $I=Ni$ and $\left<\Delta I^2\right>=N\left<\Delta i^2\right>$. If $N$ is proportional to the bulk concentration $C$ then: 
\begin{equation}\label{eqHooge}
S_1\propto \frac{I^2}{C}\times\frac{\left<\Delta i^2\right>}{i^2}
\end{equation}
For independent charge carriers $\left<\Delta i^2\right>/i^2$ is independent of $C$ and Eq.\ref{eqHooge} gives $S_1\propto C^{-1}$ (Hooge's formula). This is in contradiction with the master curve (Fig.\ref{figSI2}) obtained for KCl concentrations varying by 2 orders of magnitude (i.e. a factor much larger than the "width" of the master curve). This disagreement has been already pointed out~\cite{Powell:2009fk}, and efforts to reconcile experiment and Hooge's formula invoke ion concentrations inside the pore different from the bulk ones due to the charges of the pore wall ($N\not\propto C$). At pH~=~2 surface charges are clearly annihilated (Fig.\ref{figIV} no rectifying effect) but noise data at this pH still remain on the master curve (Fig.\ref{figSI2}). This result rules out concentration fluctuations due to pore wall charges as responsible for pink noise but also any mechanism involving individual fluctuations of ions mobility. 

On the contrary, our results for KCl solutions give evidence for cooperative effects on ions mobility. These cooperative effects are not observed in the bulk and are due to confinement. For KCl solutions they manifest themselves only on conductivity fluctuations but not on its averaged value that follows (within error bars) the expected concentration dependence. But for ionic liquids, cooperativity is even more evident. Ionic liquids are known to self-organise into liquid crystal-like structure when the side chain of the cation is long enough~\cite{Binnemans:2005} (e.g. butyl of BMIM vs. ethyl of EMIM). More recently, the phase behavior has been found to depend on the external electric field~\cite{Wang:2009ys,Xie:2010vn}. These effects should be responsible for the different conductivity properties of ionic liquids in confined geometry, i.e. facilitated transport (low noise and $\sigma_{\text{conf}}\gg \sigma_{\text{bulk}}$ for BMIM-TFSI) or jammed-like transport (high noise and $\sigma_{\text{conf}}\ll \sigma_{\text{bulk}}$ for EMIM-SCN and BMIM-methyl sulfate and methoxyethoxyethyl sulfate~\cite{Davenport:2009fk}). The noise reduction we have observed with one ionic liquid probably explains success recently reported for nanopore sensing of small molecules~\cite{Jayawardhana:2009} or DNA~\cite{Zoysa:2009} using ionic liquids and is quite promissing for future applications in this field. Even if $1/f$ noise of ionic conductance in single nanopores is not yet understood, its origin has to be searched in relation with the slow dynamics of the confined electrolyte that should display jammed like features~\cite{Biroli:2007} such as those encountered for quasi-1D transport~\cite{Nagatani:2002fk}. Finally, we think that these noise measurements should be quite interesting not only for the improvement of single molecule detection but also for the development and improvement of electrical batteries and cells that increasingly use confined geometries.

\begin{acknowledgements}
We thank Jean Le Bideau for enlightening discussions concerning ionic liquids.
\end{acknowledgements}

\end{document}